\documentclass[aps,prd,reprint,nofootinbib,superscriptaddress]{revtex4-1}

\usepackage{amsmath,amssymb,amsfonts}
\usepackage{mathrsfs}
\usepackage{bbm}
\usepackage{slashed}
\usepackage{graphicx}
\usepackage{verbatim}
\usepackage[T1]{fontenc}
\usepackage[utf8]{inputenc}
\usepackage{hyperref}
\usepackage{ifthen}
\usepackage{forloop}
\usepackage[english]{babel}
\usepackage{mathbbol}
\usepackage[usenames,dvipsnames]{xcolor}
\usepackage[normalem]{ulem}
\usepackage{tabularx, booktabs}
\usepackage{subcaption}

\newcommand{\bea}{\begin{eqnarray}}
\newcommand{\eea}{\end{eqnarray}}
\newcommand{\be}{\begin{equation}}
\newcommand{\ee}{\end{equation}}

\definecolor{darkgreen}{rgb}{0,0.4,0}

\begin{document}

\author{Mingwei Dai}
\email{mdai07@syr.edu}
\affiliation{Department of Physics, Syracuse University, Syracuse, NY 13244}
\author{Walter Freeman}
\email{wafreema@syr.edu}
\affiliation{Department of Physics, Syracuse University, Syracuse, NY 13244}
\author{Jack Laiho}
\email{jwlaiho@syr.edu}
\affiliation{Department of Physics, Syracuse University, Syracuse, NY 13244}
\author{Marc Schiffer}
\email[]{mschiffer@perimeterinstitute.ca}
\affiliation{Perimeter Institute for Theoretical Physics, 31 Caroline St. N., Waterloo, ON N2L 2Y5, Canada}
\author{Judah Unmuth-Yockey}
\email{jfunmuthyockey@gmail.com}
\affiliation{Fermi National Accelerator Laboratory, Batavia, IL 60510}

\title{An improved algorithm for dynamical triangulations and simulations of finer lattices}

\begin{abstract} 

We introduce a new algorithm for the simulation of Euclidean dynamical triangulations that mimics the Metropolis-Hastings algorithm, but where all proposed moves are accepted.  This rejection-free algorithm allows for the factorization of local and global terms in the action, a condition needed for efficient simulation of theories with global terms, while still maintaining detailed balance.  We test our algorithm on the $2d$ Ising model, and against results for EDT obtained with standard Metropolis.  Our new algorithm allows us to simulate EDT at finer lattice spacings than previously possible, and we find geometries that resemble semiclassical Euclidean de Sitter space in agreement with earlier results at coarser lattices.  The agreement between lattice data and the classical de Sitter solution continues to get better as the lattice spacing decreases.
 
\end{abstract}

\maketitle
\section{Introduction}

Lattice field theory is a versatile tool for investigating quantum field theories in regimes where perturbation theory is not applicable.  Though it is well-known that quantum gravity is non-renormalizable in perturbation theory \cite{Goroff:1985th}, there remains the possibility that there is a non-trivial ultra-violet fixed point, such that the theory is effectively renormalizable nonperturbatively. Exploring the asymptotic safety scenario for quantum gravity \cite{Weinberg:1980gg} requires non-perturbative methods such as the Functional Renormalization Group (FRG) \cite{Wetterich:1992yh, Morris:1993qb, Ellwanger:1993mw, Reuter:1996cp} (see, \cite{Dupuis:2020fhh} for a review), or lattice methods such as Euclidean/causal dynamical triangulations (EDT/CDT)\cite{Ambjorn:1991pq, Agishtein:1991cv, Catterall:1994pg, Ambjorn:2005db, Ambjorn:2005qt, Laiho:2016nlp} (see \cite{Loll:2019rdj, Ambjorn:2022naa} for reviews). 

In this paper we focus on lattice methods, where the emergence of semiclassical geometries from such calculations is by no means guaranteed.  Numerical EDT calculations have nonetheless demonstrated the recovery of geometries resembling semiclassical de Sitter space in dimension four, and the agreement between the lattice theory and the classical de Sitter solution improves as the lattice spacing is made finer \cite{Laiho:2016nlp}. Pushing these studies to still finer lattice spacings is important for testing the asymptotic safety scenario, but it is difficult in current implementations of lattice gravity using standard algorithms.  In this work we introduce and test an algorithm for EDT that leads to significant speed-ups over current algorithms, particularly at finer lattice spacings, which are necessary to thoroughly investigate the phase diagram and the emergent classical geometry.

Numerical calculations in lattice field theory typically use Markov-chain Monte Carlo methods to stochastically sample the phase space and estimate the path integral, generating a series of configurations (for instance, of gauge fields) that is representative of the equilibrium character of the system. Since this is a stochastic sampling, more accurate results can be achieved by running the calculation for more time. However, lattice discretizations also involve approximations: finite volumes, discretization of spacetime, and often others. These approximations contribute to the error budget, and these errors can generally be reduced by increasing the lattice volume and decreasing the lattice spacing -- both of which increase the computational cost. Thus, lattice calculations are notoriously hungry for faster algorithms and faster computers, since they can be used to reduce either finite volume/discretization errors or statistical errors.

The contributions to the numerical cost of a simulation vary depending on the problem at hand. For instance, in many cases we want to simulate close to a continuous phase transition, for example in Monte Carlo simulations of the Ising model; here the main cost for a careful characterization of the phase transition is due to the critical slowdown \cite{RevModPhys.49.435}. In lattice quantum chromodynamics (QCD) this is also true; the inclusion of matter fields, especially fermions, results in extra computational complexity. The most general approach to Markov-chain Monte Carlo is the famous Metropolis-Hastings algorithm \cite{10.1063/1.1699114}, which often serves as a ``fallback'' approach for simulations where no other technique is known. In lattice QCD, we can greatly accelerate the progress through phase space using global update techniques such as hybrid Monte Carlo (HMC) \cite{Duane:1987de}, allowing for a large change in the configuration that nonetheless results in a small change in the action.  For the Ising model there exist the even more efficient cluster algorithms \cite{PhysRevLett.62.361} (see \cite{Clark:2006wq} for an overview), which also make large changes in the configuration with high acceptance, leading to fast sampling of the phase space.

Lattice quantum gravity calculations attempt to evaluate the path integral over spacetime geometries using a similar Markov-chain Monte Carlo calculation \cite{Bilke:1994yf}. Spacetime is discretized as a mesh of connected simplices; the connections between these simplices characterize the geometry, and the path integral becomes a sum over all possible geometries, weighted by the Boltzmann factor for the action. Since there is no known global update algorithm (such as RHMD for lattice QCD), calculations have generally used the Metropolis-Hastings algorithm. The Metropolis suggestions come from a set of local transformations known as the Pachner moves \cite{PACHNER1991129, Agishtein:1991cv, GROSS1992367, Catterall:1994sf}, which are ergodic and reversible as required. The gravity simulations suffer from the usual critical slowing down associated with approaching a phase transition, but they also suffer from dramatically lower acceptance of the local moves in the region of the phase diagram corresponding to the putative continuum limit. In this work we introduce a rejection free algorithm that addresses this second problem and leads to a speed-up of up to two orders of magnitude for fine lattice generation compared to standard Metropolis.  

There is a long history of using rejection free algorithms to evaluate the partition function of models in the low temperature regime where the acceptance of standard Metropolis becomes very small \cite{Norman1972331, Bortz197510, gillespie1976general, Gillespie19772340, gentle2003random, schulze2004hybrid, schulze2002kinetic} (see \cite{SCHULZE20082455} for an overview), but the algorithm we introduce here is new as far as we are aware, in its ability to handle a change in action with local and global terms, where the global terms arise in lattice gravity due to the fluctuations in volume of the geometry.  We note that the algorithm introduced here is more general than its application to gravity, and it may be of interest for simulating other discrete systems that involve global contributions to the action when the acceptance of standard Metropolis is very low.

Our application to gravity requires that we take account of global contributions when computing the probability of making an update move. In order to separate out the global and local contributions, we must generalize the notion of the Metropolis accept probability to be no longer piecewise defined.  We refer to this generalized weight as the {\it ponderance}, a quantity we discuss. We show that this method leads to an $\mathcal{O}(\log (N))$ scaling as a function of the lattice size $N$, independent of the Metropolis accept rate. 

While the Ising model does not feature a global contribution to the action, we use the example of the $2d$-Ising model to show that the algorithm framed in terms of ponderances generates the correct canonical ensemble when it is sampled correctly.  We also validate the new rejection free algorithm for EDT by comparing with results generated with (a parallel version \cite{Laiho:2016nlp} of) standard Metropolis.  Finally, we use our new algorithm to generate EDT ensembles that would not have been possible with our previous algorithms, and we study the phase diagram of the theory in a region in which we were previously unable to simulate.  We find that the agreement between our geometries and the classical de Sitter solution continues to improve at finer lattice spacing.

This paper is organized as follows:  In Section \ref{sec:EDT} we review the lattice gravity formulation of EDT in four dimensions. In Section \ref{sec: Alg}, after reviewing the basic idea of rejection-free algorithms, we introduce our new algorithm, and we discuss its detailed implementation.  Section \ref{sec: Tests} discusses performance and validation tests of the algorithm.  These tests include a study of the algorithm for the $2d$-Ising model and a series of tests for $4d$ EDT, the main target of this work.  Section \ref{sec:Results} presents the results of our study of EDT at finer lattice spacings than previously possible, including a study of the phase diagram in this region and a comparison of the new ensembles to the de Sitter solution.  We conclude in Section \ref{sec: Concl}.

\section{Euclidean Dynamical Triangulations}\label{sec:EDT}

In Euclidean quantum gravity the partition function is formally given by a path integral sum over all geometries
\bea\label{eq:part}  Z_E = \int {\cal D}[g] \, e^{-S[g]},
\eea
where the exact form of the action $S$ depends on the chosen approach to quantum gravity. 
    
In dynamical triangulations, the path integral is formulated directly as a sum over geometries, without the need for gauge fixing or the introduction of a metric.  The dynamical triangulations approach is based on the conjecture that the path integral for Euclidean gravity is given by the partition function \cite{Ambjorn:1991pq, Bilke:1998vj}
\bea\label{eq:Z} Z_E = \sum_T \frac{1}{C_T}\left[\prod_{j=1}^{N_2}{\cal O}(t_j)^\beta\right]e^{-S_{ER}}
\eea
where $C_T$ is a symmetry factor that divides out the number of equivalent ways of labeling the vertices in the triangulation $T$. Furthermore, $S_{ER}$ is a discretized version of the Einstein-Hilbert action
\bea  \label{eq:ERcont} S_{EH} =  -\frac{1}{16 \pi G}\int d^4x \sqrt{g} (R - 2\Lambda),
\eea
with $R$ the Ricci curvature scalar, $g$ the determinant of the metric tensor, $G$ Newton's constant, and $\Lambda$ the cosmological constant.  The term in brackets in Eq.~(\ref{eq:Z}) is a nonuniform measure term \cite{Bruegmann:1992jk}, where the product is over all two-simplices (triangles), and ${\cal O}(t_j)$ is the order of triangle $j$, i.e. the number of four-simplices to which the triangle belongs.  This corresponds in the continuum to a nonuniform weighting of the measure in Eq.~(\ref{eq:part}) by $\prod_x \sqrt{g}^{\beta}$, with $\beta$ a free parameter in the simulations.  

In four dimensions the discretized version of the Einstein-Hilbert action is the Einstein-Regge action \cite{Regge:1961px}
\begin{equation} \label{eq:GeneralEinstein-ReggeAction}
S_{ER}=-\kappa\sum_{j=1}^{N_2} V_{2}\delta_j+\lambda\sum_{j=1}^{N_4} V_{4},
\end{equation}
\noindent where $\delta_j=2\pi-{\cal O}(t_j)\arccos(1/4)$ is the deficit angle around a triangular hinge $t_j$, with ${\cal O}(t_j)$ the number of four-simplices meeting at the hinge, $\kappa=\left(8\pi G \right)^{-1}$, $\lambda=\kappa\Lambda$, and the volume of a $d$-simplex is 
\begin{equation} \label{eq:SimplexVolume}
V_{d}=\frac{\sqrt{d+1}}{d!\sqrt{2^{d}}}a_{\rm lat}^d,
\end{equation}
\noindent where the equilateral $d$-simplex has a side of length $a_{\rm lat}$.  After performing the sums in Eq.~(\ref{eq:GeneralEinstein-ReggeAction}) one finds
\begin{equation}\label{eq:DiscAction}
S_{ER}=-\frac{\sqrt{3}}{2}\pi\kappa N_{2}+N_{4}\left(\kappa\frac{5\sqrt{3}}{2}\mbox{arccos}\frac{1}{4}+\frac{\sqrt{5}}{96}\lambda\right),
\end{equation}
where $N_i$ is the number of simplices of dimension $i$.  We rewrite the Einstein-Regge action in the simple form 
\bea\label{eq:ER}  S_{ER}=-\kappa_2 N_2+\kappa_4N_4,
\eea
where we introduce the parameters $\kappa_4$ and $\kappa_2$ in place of $\kappa$ and $\lambda$ for convenience in the numerical simulations. Hence, the Einstein-Regge action depends only on global quantities of the discretized geometry, namely the number of four- and two-simplices, but not on any local structure.  The local structure does enter through the measure term, however.

Geometries are constructed by gluing together four-simplices along their ($4-1$)-dimensional faces.  The four-simplices are equilateral, with constant edge length $a_{\rm lat}$.  The set of all four-geometries is approximated by gluing together four-simplices, and the dynamics is encoded in the connectivity of the simplices.  Most early simulations of EDT \cite{Ambjorn:1991pq, Agishtein:1991cv, Catterall:1994pg, deBakker:1994zf, Ambjorn:1995dj, Egawa:1996fu} used a set of triangulations that satisfies the combinatorial manifold constraints, so that each distinct $4$-simplex has a unique set of 4+1 vertex labels.  The combinatorial manifold constraints can be relaxed to include a larger set of degenerate triangulations in which distinct four-simplices may share the same 4+1 (distinct) vertex labels \cite{Bilke:1998bn}.  It was shown that the finite size effects of degenerate triangulations are a factor of $\sim$10 smaller than those of combinatorial triangulations \cite{Bilke:1998bn}.  This makes it easier to study the phase diagram with degenerate triangulations, although there appears to be no essential difference in the phase diagram between degenerate and combinatorial triangulations in four dimensions \cite{Ambjorn:2013eha, Coumbe:2014nea, Asaduzzaman:2022kxz}.

It is necessary to constrain the four-volume of the lattices, either to a fixed value or to a window of values, in order to efficiently sample the path integral at large volumes.  We choose to simulate at a fixed fiducial number of four-simplices $N_4^f$, though in practice the lattice volume fluctuates somewhat around this value.  This is due to the fact that the local Pachner moves are not ergodic for fixed volume, so the volume must be allowed to fluctuate in order to properly sample the path integral.  To keep the volume fluctuations under control, a volume preserving term $\delta\lambda|N_4^f - N_4|$ is added to the action to keep the volume close to the fiducial number of four-simplices $N_4^f$.  Although the simulations are only ergodic in the limit that $\delta\lambda$ goes to zero, for sufficiently small $\delta\lambda$ this is expected to introduce only a small systematic effect.  In practice we have found that varying the value of a small but non-zero $\delta\lambda$ has little effect on the observables obtained from the simulations, as discussed in more detail in \autoref{sec:EDTvalid}.  Note that the presence of fluctuating global terms in the action such as this one complicates the implementation of standard rejection free algorithms.

\section{A new rejection-free algorithm}\label{sec: Alg}

For EDT, the Metropolis accept rate can be quite low, and it decreases substantially as one approaches the continuum limit; this provides a critical limitation on the ability to make contact with the continuum, since a computer implementing the Metropolis algorithm will spend most of its time proposing moves that it will then reject.

We thus propose a {\it rejection-free} update that cannot fail to make an update. This algorithm offers significant performance gains for finer lattices since it overcomes the low acceptance rate of traditional Metropolis. Rejection-free approaches to Markov-chain Monte Carlo have been developed previously, although to our knowledge not for dynamical triangulations. 

The algorithm we present generates a Markov chain based on the Pachner moves and, despite the presence of a global contribution to the action, exhibits a na\"{i}ve $\mathcal{O}(\log(N))$ scaling of the cost per move. It is not limited to dynamical triangulations and can be applied to other discrete systems; we use the $2d$ Ising model as a test case.

In this section we first briefly review the general idea of rejection free algorithms for local actions, and then introduce an alternative mathematical formulation of Markov chain Monte Carlo to efficiently employ rejection free algorithms for actions with global contributions; this requires an alternative to the Metropolis accept probabilities that we call \textit{ponderances}.

\subsection{Rejection free algorithm for local actions}
\label{sec:RFlocal}

The Metropolis algorithm generates a Markov chain that progresses through configuration space by proposing moves until one is accepted. As discussed previously, this becomes slow if the accept rate is low. We wish to determine which move $i$ will be the next one to be accepted in a faster way.

We call the Metropolis accept probability of a move at site $i$ $P(i)$.  Then, the probability $\tilde{P}(i)$ that the move at any given site $i$ will be the one eventually accepted is

\begin{equation}
\tilde{P}(i) = \frac{P(i)}{\sum_j P(j)},
\end{equation}
that is, the probability of eventually accepting the move at any particular site $i$ is a fraction of the
sum over Metropolis probabilities for all possible successive moves. 

The essence of the rejection-free algorithm (for purely local actions) is to instead calculate in advance the Metropolis accept probabilities of all possible moves, then determine the next move that the Metropolis algorithm would eventually accept.

Once this move is made, the new accept probabilities of each possible move could in principle change. However, for many models, the accept probability depends only on the local neighborhood of a proposed move. For instance, in the $2d$ Ising model, flipping one spin will change the accept probabilities of five moves: itself and its four neighbors.

In particular, a standard rejection free algorithm for a lattice with $N$ sites and a single move that can be performed (such as in the Ising model), follows the steps:

\begin{enumerate}
    \item Initialize
    \begin{itemize}
        \item Save Metropolis probabilities $P(i)$ to perform a move at lattice site $i$
    \end{itemize}
    \item Find and perform next move
    \begin{itemize}
        \item Generate a random number $r\in(0, \sum_i P(i)]$ 
        \item Determine the site $j$  that corresponds to the random number $r$\footnote{When saving the summed probabilities $\hat{P}(j)=\sum_{i\leq j}P(i)$ in a list, this corresponds to choosing the site $j$ with $\hat{P}(j-1)<r\leq\hat{P}(j)$, with $\hat{P}(0)=0$. We discuss a more efficient implementation via binary trees in \autoref{sec:implementprob}}
        \item Perform the move at the chosen site $j$
    \end{itemize}
    \item Update the stored Metropolis probabilities 
    \begin{itemize}
        \item Determine the sites $k$ for which the accept probabilities have changed after making the move at site $j$ 
        \item Update or recompute the Metropolis probability of accepting the move at each site $k$
    \end{itemize}
    \item Repeat from step $2)$
\end{enumerate}

We can already see the potential speed-up: if the cost of step $3)$ is fast compared to the cost of proposing a large number of rejected moves before one move is accepted, this algorithm will be faster. This will generally be true if the number of moves whose probabilities need to be updated is less than the number of rejected moves per accepted move in standard Metropolis.

In the Metropolis algorithm, the Markov chain often includes multiple steps in succession where the lattice does not change due to a series of proposed moves being rejected.  To statistically reproduce the behavior of the Metropolis algorithm using the rejection-free approach, we need to determine the number $n_{\rm {rej}}$ of moves that would have been rejected before finally accepting one. As we show in appendix \ref{app:nreject}, 
\begin{equation}
     n_{\text{rej}} = \text{floor}(\log_{1-P(i)/N}(\eta))
\label{eq:nreject}
\end{equation}
with the total lattice size $N$ and a random number $\eta$ such that
\begin{equation}
    \eta\in(0,1)\,.
\end{equation}
We note that this relation was derived for the rejection-free algorithm assuming that Metropolis probabilities are being used to decide which move is accepted next (see steps 1 and 3 above). For applications with a global contribution to the action such as EDT a different approach is needed, as we show in \autoref{sec:globact}.  In that case $n_{\rm{rej}}$ in \eqref{eq:nreject} can no longer be used to compute the number of moves a Metropolis algorithm would have rejected.

\subsection{Evaluating the algorithmic complexity of the rejection-free algorithm}
\label{sec:complexity}
Suppose that we wish to implement the rejection-free algorithm on a model with $N_m$ possible moves, and that after making a move, the accept probabilities of $N_c$ other moves will change. 

The algorithmic complexity of each accepted move is thus:

\begin{enumerate}
    \item $\mathcal O(\log N_m)$ to select the move to make (using a binary decision tree; see \autoref{app:bintrees})
    \item $\mathcal O(1)$ to actually make the move
    \item $\mathcal O(N_c)$ to calculate the new accept probabilities of each move whose accept probability has changed; this step can be quite readily parallelized
    \item $\mathcal O(N_c \log N_m)$ to update the binary decision tree with those changed probabilities; this step can also be parallelized
\end{enumerate}

In general, we expect $N_c \ll N_m$ if the action is local; regardless of the size of the lattice, only a small number of probabilities must be updated once a move is made. This algorithm thus does not include any steps of $\mathcal O(N_m)$ and does not depend on the Metropolis accept probability, so it does not slow down for large lattices or low accept rates.

\subsection{Incorporating global changes to the action}
\label{sec:globact}

For actions that are not local, it may be the case that $N_c \propto N_m$, leading to high computational cost for steps (3) and (4) above. In EDT simulations, there are global contributions to the action that would require recomputing all Metropolis probabilities after each local move. The Einstein-Regge action \eqref{eq:ER} only depends on the total number of two- and four-simplices, and not on any local structure. However, the simulations do depend on the local geometry through the measure term in \eqref{eq:Z}. Exponentiating the measure term and including it as a contribution to the action, we write schematically
\begin{equation}
S_{\mathrm{tot}}=S_{\mathrm{loc}}+S_{\mathrm{glob}},
\end{equation}
where the Einstein-Regge action is part of the global action $S_{\mathrm{glob}}$, and the measure term is part of the local action $S_{\mathrm{loc}}$. The volume preserving term also contributes to the global part of the action, as it only depends on the number of four-simplices. 

Since the total number of simplices can change after each move, these global contributions to the action cause issues with the above algorithm. They are uniform for each Pachner move type, but the global action of each move type can change every time a move is made. This would in principle require updating the accept probability of every move on the lattice, requiring a very large number of updates of the probability decision trees. This would cause an unacceptable impact on performance.

Since the global contributions to the change in action $\Delta S_{\rm glob}$ are the same for every move type, one might imagine separating out the contribution to the accept probability from the local and the global terms in the action. In this approach, one could imagine a rejection-free algorithm that stores only the local contribution to the accept probability. Then one could choose the next Pachner move in a two-step process:

\begin{enumerate}
\item Determine which of the five types of Pachner move will be made, using only the global factors and the summed accept probabilities of all moves of each type.
\item Once the Pachner move type is chosen, then use the local contributions to the accept probability to choose which simplex of the appropriate type to apply that Pachner move to.
\end{enumerate}

If the accept probability factorized neatly into local and global contributions, this approach would work. However, in the conventional Metropolis algorithm, it does not. Recall the definition of the Metropolis accept probability:
\begin{equation}
\label{eq:Metacc}
P(A \rightarrow B)=
    \begin{cases}
        1 & \text{if }  S_B <  S_A\\
        \exp( S_A - S_B) & \text{if }  S_B >  S_A
    \end{cases}
\end{equation}

The piecewise definition of the accept probability thus means the local and global parts do not factorize. Consider a move that reduces the local action substantially; regardless of the global factors, its accept probability will be 1. Thus, changes in the global factors for one Pachner move type will affect the Metropolis probabilities of some moves but not others.

This piecewise definition leads to a situation where
\begin{equation}
P(A \rightarrow B)\neq \left(P(A \rightarrow B)\right)_{\mathrm{loc}}\,\left(P(A \rightarrow B)\right)_{\mathrm{glob}}\,.
\end{equation}

Since the conventional Metropolis accept probability does not factorize neatly into local and global parts, we must either update every accept probability whenever the global factors change (which is computationally expensive) or develop an alternative to the Metropolis accept probability that {\it can} be factorized into local and global contributions. 

\subsection{An alternative to the accept probability: the ``ponderance''}

The rejection-free algorithm presented in \autoref{sec:RFlocal} is designed to reproduce the same Markov chain as the Metropolis algorithm, and thus uses the Metropolis accept probabilities to calculate the next move that would be accepted and the number of moves that would be rejected before this accepted move. However, the notion of ``accept probability'' here is purely vestigial: nothing is being accepted or rejected, and these values do not represent the probability of any event. Instead, the value stored in the decision trees represents the \textit{relative} likelihood that a move will be chosen as the next move in the Markov chain. Thus, we are free to store any other quantity in the decision trees so long as the relative likelihoods are in the correct proportion, ensuring that the algorithm preserves detailed balance.

Detailed balance
\begin{equation}
\label{eq:detbal}
    \frac{P(A \rightarrow B)}{P(B \rightarrow A)}=\frac{P(B)}{P(A)}\,,
\end{equation}
states that the ratio between the probability of the transition from state A to state B ($P(A \rightarrow B)$) and the probability of the reverse transition from state B to state A ($P(A \rightarrow B)$) is equal to the ratio between the probability of state B being sampled ($P(B)$) and the probability of state A being sampled ($P(A)$). Since algorithms like these are generally applied to the canonical ensemble, the target distribution is $P(X) \propto e^{-S_X}$. The Metropolis algorithm preserves detailed balance using the definition in eq. \ref{eq:Metacc}, since \begin{equation}
 \frac{P(A \rightarrow B)}{P(B \rightarrow A)} = \frac{e^{(S_A-S_B)}}{1}=\frac{e^{-S_B}}{e^{-S_A}}.
\end{equation}
But this definition is not the only one that maintains detailed balance. As an alternative we define a new quantity, the {\it ponderance} $\mathcal P(A \rightarrow B)$, which we use to replace the piecewise-defined Metropolis probability from \eqref{eq:Metacc} in our algorithm:
\begin{equation}
\label{eq:Ponddef}
\mathcal{P}(A\rightarrow B)=e^{\frac{1}{2}\left(S_A-S_B\right)}\,.
\end{equation}
This quantity cannot be interpreted as a Metropolis accept probability, since it may be greater than unity. However, in the rejection-free algorithm, the Metropolis accept probability is repurposed as a selection weight in a decision tree; since it is not actually the probability of any event, there is no mathematical reason that it cannot be greater than unity.  The factor $1/2$ in \eqref{eq:Ponddef} is necessary to maintain detailed balance, as we discuss later.

Because it acts as a selection weight in our algorithm, we could appropriately refer to this quantity $e^{\frac{1}{2}\left(S_A-S_B\right)}$ as the {\it weight}. However, this term in Markov-chain Monte Carlo is usually reserved for sampling weights once the sequence of configurations is generated (such as in the technique called ``reweighting''), and we use it with this meaning below in \eqref{eq:weightdef}. To avoid confusion, we introduce a new term for it.

The term ``ponderance'' for $e^{\frac{1}{2}\left(S_A-S_B\right)}$ is a mathematical neologism: we  choose this word since its first letter $\mathcal P$ evokes its origin as an alternative to the Metropolis accept probability $P$, while its meaning of ``weight'' in archaic English describes its role in our algorithm as a selection weight.

It is not piecewise defined, so it cleanly separates into local and global contributions:
\begin{equation}
\begin{aligned}
\mathcal{P}(A\rightarrow B)=&e^{\frac{1}{2}\left(S_{A\,,\mathrm{loc}}+S_{A\,,\mathrm{glob}}-(S_{B\,,\mathrm{loc}}+S_{B\,,\mathrm{glob}})\right)}\\
=& e^{\frac{1}{2}\left(S_{A\,,\mathrm{loc}}-S_{B\,,\mathrm{loc}}\right)}\,e^{\frac{1}{2}\left(S_{A\,,\mathrm{glob}}-S_{B\,,\mathrm{glob}}\right)}\\
=&\left(\mathcal{P}(A\rightarrow B)\right)_{\mathrm{loc}}\,\left(\mathcal{P}(A\rightarrow B)\right)_{\mathrm{glob}}\,.
\end{aligned}
\end{equation}

Even though the ponderance of a move cannot be interpreted as a probability directly, it is proportional to the probability that it will be chosen as the next move in the Markov chain. Thus, replacing probabilities by ponderances in the implementation discussed in \autoref{sec:implementprob} retains the character of the Metropolis Markov chain.

Once again we need to make sure to reproduce the correct importance sampling, emulating the behavior of the Metropolis Markov chain whereby configurations are repeated whenever a move is rejected. The purely local rejection-free algorithm based on accept probabilities that is designed to exactly mimic the Metropolis algorithm requires a determination of the number of rejected moves $n_{\rm {rej}}$ (eq.\ref{eq:nreject}) at each step. We must derive an equivalent for our new algorithm that instead uses ponderances.

The transition probability from state A to B can be calculated as the probability of the particular move being chosen multiplied by the probability the move is accepted, and since this is a rejection-free algorithm, $P( A\rightarrow B _\mathrm{accepted}) = 1$ for all possible moves, so the transition probability is just the probability of the move being chosen. We choose the move by generating a random number between $0$ and the total ponderance of all possible moves in state A $\left(\sum_{i=1}^{N}\mathcal{P}(A\rightarrow i)\right)$. Therefore, the transition probability of move $A\rightarrow B$ is:
\begin{equation}
\label{eq:transitionP}
\begin{aligned}
P(A\rightarrow B)= &P(A\rightarrow B _\mathrm{chosen})\times 1 \\=& \frac{\mathcal{P}(A\rightarrow B)}{\sum_{i=1}^{N}\mathcal{P}(A\rightarrow i)}
\end{aligned}
\end{equation}
Plugging \eqref{eq:transitionP} into \eqref{eq:detbal}, using \eqref{eq:Ponddef}, we get:
\begin{equation}
    \frac{P(B)}{P(A)} = \frac{e^{-S_B}\sum_{i=1}^{N}\mathcal{P}(B\rightarrow i)}{e^{-S_A}\sum_{i=1}^{N}\mathcal{P}(A\rightarrow i)}
\end{equation}
This means that if we sample after a fixed number of moves, the distribution we get is $P(A) \propto e^{-S_A}\sum_{i=1}^{N}\mathcal{P}(A\rightarrow i)$, which deviates from our target distribution. Hence, this version does not satisfy detailed balance; we will call this the \textit{move accumulating} version (see \autoref{sec:validandtest}). 

However, we could compensate for this deviation by assigning a weight $\omega$
\begin{equation}
    \omega=\frac{1}{\sum_{i=1}^{N}\mathcal{P}(A\rightarrow i)}\,
    \label{eq:weightdef}
\end{equation}
to each state; this weight is the inverse of the total ponderance of moves exiting that state. We then sample after a fixed amount of accumulated weight. This has the effect of making it more likely to sample states with a large weight, or equivalently a small total ponderance. This is analogous in the Metropolis algorithm to the Markov chain getting ``stuck'' in a state when all moves away from that state have low accept probabilities, for example a highly magnetized state at low temperature in the $2d$ Ising model. The weight $\omega$ plays a similar role in our algorithm to the quantity $n_{\rm{rej}}$ that is used in rejection-free algorithms based on Metropolis probabilities (See \eqref{eq:nreject}).

We can also understand the role of the weight factor at low ponderances: low ponderances correspond to moves that are unlikely to happen in the Metropolis algorithm. Hence, a state with a small average ponderance of moves leading away from it will remain unchanged for a long Monte Carlo time. Conversely, such a state will have a large weight, such that the weight of a move can be understood as the dwell time of the state before the move.

We might imagine that this algorithm generates a weighted ensemble, producing a chain of configurations combined with a weight factor for each; then, measuring any observable on this ensemble involves computing a weighted average. This is commonly done in lattice QCD calculations using ``reweighting'', but it involves an extra complication in analysis. Instead, we choose to sample the Markov chain at fixed intervals of accumulated weight factor $\omega$, such that configurations with large weight factor are more likely to be sampled but all sampled configurations contribute equally in analysis; we call this the \textit{weight accumulating} version (see \autoref{sec:validandtest}). This reproduces the importance sampling of the Metropolis algorithm, as we demonstrate in \autoref{sec:validandtest}, without the complication of the weighted ensemble.

\subsection{Implementation details: binary trees and parallelization}
\label{sec:implementprob}

In this subsection we label moves, rather than lattice sites where the move can be performed.
In order to implement this procedure, we must find an efficient way to select the next move $i$ from among all possible moves such that the probability of choosing move $i$ is proportional to $\mathcal{P}(i)$. Since the total number of possible moves may be quite large, we want to avoid any operations that require us to iterate over all possible moves.

We achieve this by storing the local part of the ponderances $\left(\mathcal{P}(i)\right)_{\mathrm{loc}}$ of each move in a binary decision tree. 
This procedure allows the algorithm to select the move $j$ in $\mathcal O(\log(N))$ time. In practice, for EDT simulations this is a trivial contribution to the overall runtime of the calculation.

Once the move $j$ is chosen, we must update things to reflect that move:

\begin{enumerate}
\item Actually make that move (change the geometry of a dynamical triangulation simulation, flip a spin in the Ising model, etc.)
\item Create a list of other moves whose local ponderance $\mathcal P_{\rm {loc}}$ might be changed by that move (i.e. those in the local neighborhood of $j$)
\item Recompute the local ponderances $\mathcal P_{\rm {loc}}$ of moves on that list 
\item Update the binary decision tree to reflect the set of changed local ponderances $\{\mathcal P_{\rm{loc}}\}$
\end{enumerate}

It is worth noting at this point a peculiarity of dynamical triangulation simulations. Unlike the Ising model, where there are always a fixed number of possible moves (one per spin), the collection of simplices in a dynamical triangulation simulation must grow and shrink. As simplices are created and destroyed by the Pachner moves, it is possible to dynamically prune and grow the probability tree, taking care to ensure that it stays balanced. However, this is not necessary. If a simplex is deleted from the geometry, it suffices to set $P_{\rm here}$ to zero for that move in the probability tree, ensuring that it cannot be chosen; the tree-traversal algorithm will then skip over it. Thus, in practice we create a tree large enough to hold moves for the maximum possible number of simplices, recognizing that as many as half of them may have $P_{\rm here}=0$ if those simplices currently do not exist. This may seem inefficient, but due to the $\mathcal O(\log N)$ cost of tree traversal it adds only a small fraction to the time required to find the move $j$. This allows us to keep the tree geometry fixed during a run.

\section{Performance tests and validation}\label{sec: Tests}
\label{sec:validandtest}
\subsection{Validation tests for the $2d$ Ising model}

The Ising model serves as a test bed for new Monte Carlo algorithms. In this work we focus on the $d=2$ Ising model, which is exactly solvable in the infinite-volume limit \cite{Onsager1944}. This exact solution allows analytical access to some observables also for finite systems. Hence, the $2d$ Ising model serves as a proof-of-principle demonstration that the use of ponderances in our rejection-free algorithm leads to the correct importance sampling. We test our algorithm by investigating two observables: the average magnetization as a function of the temperature $T$, and the probability distribution for the energy of a state at fixed $T$. We perform all our studies on an $N=32 \times 32$ square lattice.\\

In order to illustrate our rejection-free algorithm we compare the case where we take measurements after a certain amount of weight has accumulated (the weight accumulating version), which should mimic the counting of rejected moves for the Metropolis algorithm, and satisfies detailed balance, with the case where we take measurements after a fixed number of accepted moves (the move accumulating version), which we expect leads to the wrong importance sampling, as it does not satisfy detailed balance. This is demonstrated to be the case in our Ising model simulations. 

\subsubsection{Magnetization as a function of the temperature}
\label{sec:MofTIsing}
We first consider an analysis of the average magnetization of the system. On the lattice the magnetization is simply given by the sum of spins over all lattice sites, i.e.,
\begin{equation}
    M=\sum_{i=1}^{N} s_i\,,
\end{equation}
where $s_i=\pm1$ is the spin at lattice site $i$.\\

We test the new algorithm for temperatures $T\in [0.5,4]$, i.e., at temperatures on both sides of the transition, by comparing the results of our rejection-free algorithm with that of the standard Metropolis algorithm. At low temperatures we also compare the results of our rejection-free algorithm to a perturbative expansion of the partition function.  The perturbative expansion proceeds as follows: one counts the magnetization and the multiplicity of states at a given fixed energy
\begin{equation}
    E=-\sum_{i,j}s_i\,s_j\,.
\end{equation}
In particular, the energy above the ground state is determined by the number of misaligned boundaries, i.e., the boundary between a spin up and a spin down, on a given lattice. The ground state is the state where all spins are aligned, and it comes with multiplicity $\mathcal{N}=1$.
The leading-order perturbation to the ground state is a state with one misaligned spin, corresponding to four misaligned boundaries, and it comes with multiplicity $\mathcal{N}=N$. The next-to-leading order term corresponds to six misaligned boundaries, and so on. Within this expansion, the average of the squared magnetization reads
\begin{equation}
    \left\langle M^2 \right\rangle =\frac{\sum_{i=0} \sum_j M^2_{ij} \,\mathcal{N}_{ij} \,e^{-E_i}}{\sum_{i=0}\mathcal{N}_i \,e^{-E_i}}\,,
\end{equation}
where $E_i$ is the energy of a given state, $M^2_{ij}$ are the different squared magnetizations of a state with energy $E_i$ and $\mathcal{N}_{ij}$ are the corresponding multiplicities.  Up to ten misaligned boundaries, this ratio of sums evaluates to
\begin{widetext}
\begin{equation}
    \left\langle M^2 \right\rangle\approx\frac{2 N e^{20/T}+2 (N-2)^2 e^{12/T}+4 (N-4)^2 e^{8/T}+ \left(N^3+N^2-120 N+480\right) e^{4/T}+4 \left(N^3-7 N^2-92 N+688\right)}{N \left(2 e^{20/T}+2 N e^{12/T}+4 N e^{8/T}+(N+9) N e^{4/T}+4 (N+5) N\right)}\,.
    \label{eq:pertM}
\end{equation}
\end{widetext}
Additionally, we know that the leading order contribution for 12 misaligned boundaries comes from a state with three isolated flipped spins, with contribution
\begin{equation}
    \left\langle M^2 \right\rangle_{\text{LO}}\sim \frac{N(N-5)(N-10)}{2}\,e^{2\frac{N-12}{T}}\,,
\end{equation}
which allows us to estimate the temperature $T_{\mathrm{cut}}$ where the expansion breaks down, i.e., where
\begin{equation}
    \frac{\left\langle M^2 \right\rangle_{\text{LO}}}{\left\langle M^2 \right\rangle} \approx 0.1\,.
\end{equation} For our order of approximation, we expect this to be
\begin{equation}
    T_{\mathrm{cut}}\approx 1.1\, .
\end{equation}

\begin{figure*}
    \centering
    \includegraphics[width=.8\linewidth]{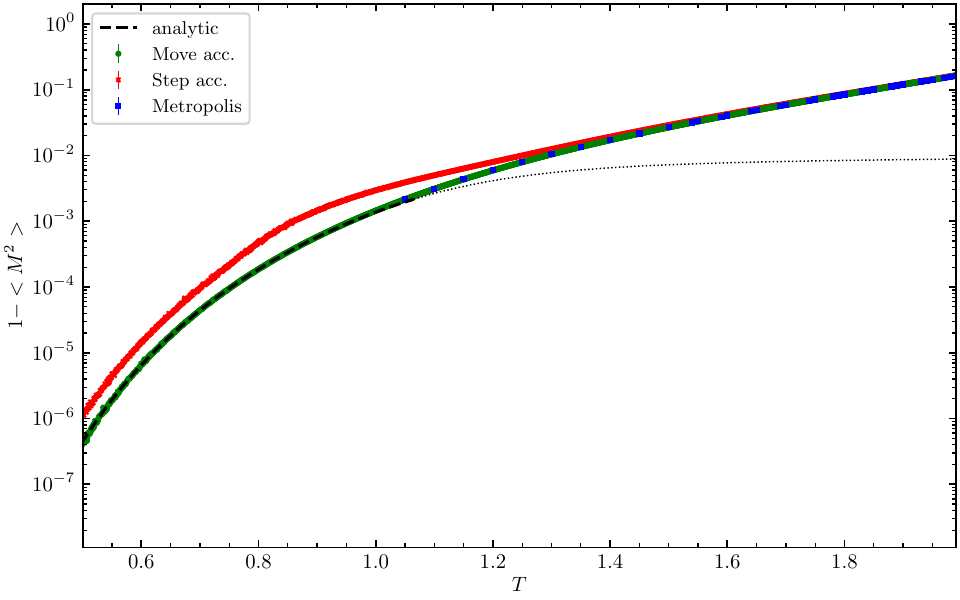}
    \caption{Temperature dependence of the squared magnetization at low temperatures. The dashed (dotted) black line shows the perturbative expansion \eqref{eq:pertM} below (above) its expected range of validity. The blue boxes (red crosses, green circles) indicate the numerical results obtained with the Metropolis algorithm (the rejection-free algorithm, measuring after a fixed number of accepted moves, the rejection-free algorithm, measuring after a fixed number of accumulated weight).}
    \label{fig:MlowTIsing}
\end{figure*}

\autoref{fig:MlowTIsing} compares the numerical results for the squared magnetization at low temperatures, showing the deviation from a completely magnetized state. We compare the results for the move accumulating and weight accumulating versions of the rejection free algorithm with that of the Metropolis algorithm. Additionally, we plot the perturbative result, Eq.~\eqref{eq:pertM}. We can clearly see that the weight accumulating version of the rejection-free algorithm is in excellent agreement with the Metropolis algorithm and with the perturbative expansion within its range of validity. However, the move accumulating version of the rejection free algorithm consistently underestimates the correct magnetization of the system and leads to wrong results. This is expected, as this version of the algorithm does not satisfy detailed balance, and hence does not perfom the correct importance sampling. 

\subsubsection{Energy distribution of states at fixed temperature}
As a second test of the validity of our rejection-free algorithm using the Ising model, we investigate the energy distribution of states at a fixed temperature. In particular, we follow \cite{PhysRevLett.76.78}, and study the probability distribution $P_k$ of finding an equilibrium state with energy $4\,k\,J$ above the ground state.  This probability is given by
\begin{equation}
    P_k(T)=\frac{g_k\,e^{-\frac{4\,k\,J}{T}}}{\sum_{k=0}^{N} g_k\,e^{-\frac{4\,k\,J}{T}}}\,,
\end{equation}
where $g_k$ is the number of possible configurations, i.e., the multiplicity of a state with energy $4\,k\,J$ above the ground state, which can be computed for a general square lattice of size $N$ \cite{PhysRevLett.76.78}. 

\begin{figure*}
    \centering
    \includegraphics[width=1\linewidth]{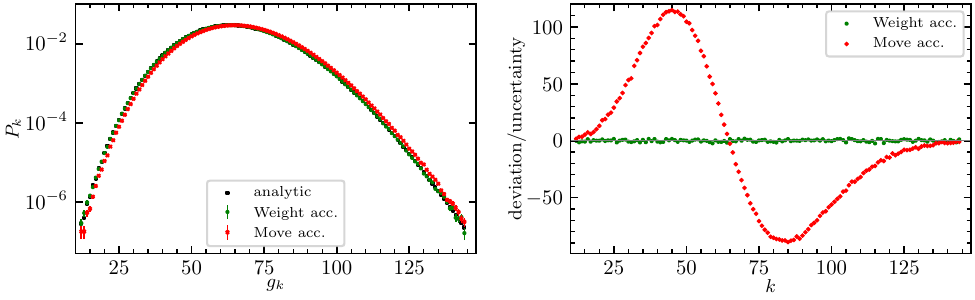}
    \caption{Left panel: probability distribution of energy states at $T=2$ for the analytical result (black squares), the weight accumulating rejection-free code (green circles), and the step accumulating rejection-free algorithm. The state $k$ corresponds to an energy of $4\,k\,J$ above the ground state.\newline
    Right panel: deviation of the numerical result from the analytical result, normalized to the statistical uncertainty of the numerical result.}
    \label{fig:EdistT2Ising}
\end{figure*}

In the left panel of \autoref{fig:EdistT2Ising} we show the analytical result for the energy distribution of an $N=32\times32$ square lattice at $T=2$ (black boxes). We compare the analytical result with that of the rejection-free algorithm. Again, we see that the move accumulating version (red crosses) gives wrong results as expected, overestimating more energetic configurations, and underestimating lower excitations to the ground state. However, the weight accumulating version of our algorithm (green dots) shows good agreement with the analytical distribution. In the right panel of \autoref{fig:EdistT2Ising} we show the deviation of the numerical results from the analytical result in units of the statistical uncertainty. While the step-accumulating version systematically deviates from the analytical solution, the weight-accumulating method shows small random deviations only. In particular, for the 133 non-vanishing values shown in \autoref{fig:EdistT2Ising}, we obtain $\chi^2_{\mathrm{Weight\,acc.}}=129$, and $\chi^2_{\mathrm{Step\,acc.}}=468353$.  

These results allow us to draw the following conclusion:  our rejection free algorithm yields the correct importance sampling in Monte Carlo simulations when properly implemented.  These Ising model calculations demonstrate that it is necessary to use the weight, \eqref{eq:weightdef}, as a proxy for Monte-Carlo time, and not the number of accepted moves. The latter results in an importance sampling which deviates systematically from the analytical results.  Our rejection-free algorithm thus passes a crucial consistency test.

\subsubsection{Performance gains}
\begin{figure}
    \centering
    \includegraphics[width=\linewidth]{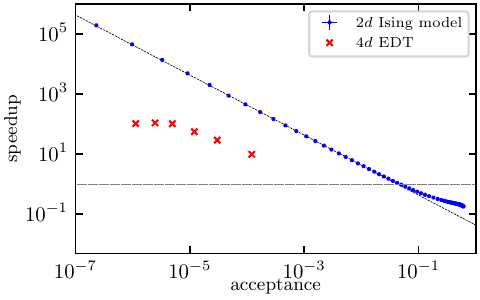}
    \caption{Performance gain of rejection-free algorithm compared to the Metropolis algorithm. We compare the average time it takes to perform a single move. For the Ising model, we tune the acceptance rate by adjusting the temperature $T$. For EDT the acceptance rate varies as a function of $\kappa_2$ and $\beta$.}
    \label{fig:IsingPerformance}
\end{figure}

For the $2d$ Ising model with nearest-neighbor interaction the performance gains of the rejection free algorithm are easy to estimate: for each accepted move of the rejection free algorithm, the probabilities or ponderances of five lattice sites have to be re-calculated. As the lattice is static, this holds for any acceptance rate, which can be adjusted by changing the temperature $T$. \autoref{fig:IsingPerformance} shows the speed-up of the rejection free algorithm compared to the Metropolis algorithm. Here, we define the speed-up as
\begin{equation}
    \text{speed-up}= \frac{\left(\text{accepted moves/second}\right)_{\text{rejection-free}}}{\left(\text{accepted moves/second}\right)_{\text{Metropolis}}}\,.
    \label{eq:speedup}
\end{equation}
As expected, the speed-up is a powerlaw as a function of the acceptance rate for small enough acceptance rates. We can also see that for relatively large acceptance rates, the Metropolis algorithm is faster than the rejection-free algorithm.  This is expected, since the advantage of rejection free disappears when the acceptance of standard Metropolis is sufficiently high, given the extra overhead of updating the ponderances in rejection free.

\subsection{$4d$ EDT}
\label{sec:EDTvalid}

\subsubsection{Validation tests}
Since our main purpose for the rejection-free algorithm is the application to EDT, we demonstrate that it produces correct results also for this case. In contrast to the $2d$ Ising model, we do not have analytical results to compare to, but we are still able to compare against results from the Metropolis algorithm. We perform two independent tests: first, we compare the determination of our preferred order-parameter for studying the phase transition in EDT between old and new algorithms for two volumes at fixed lattice spacing.  Second, we compare two lattice quantities, the average local curvature, and the tuned value of $\kappa_4$ at (approximately) fixed volume, in the limit that the volume preserving term in the action is taken to zero.\\
\paragraph{Peak height of shelling function\\[6pt]}

\begin{figure*}
    \centering
    \begin{minipage}{.495\linewidth}
    \includegraphics[width=1\linewidth]{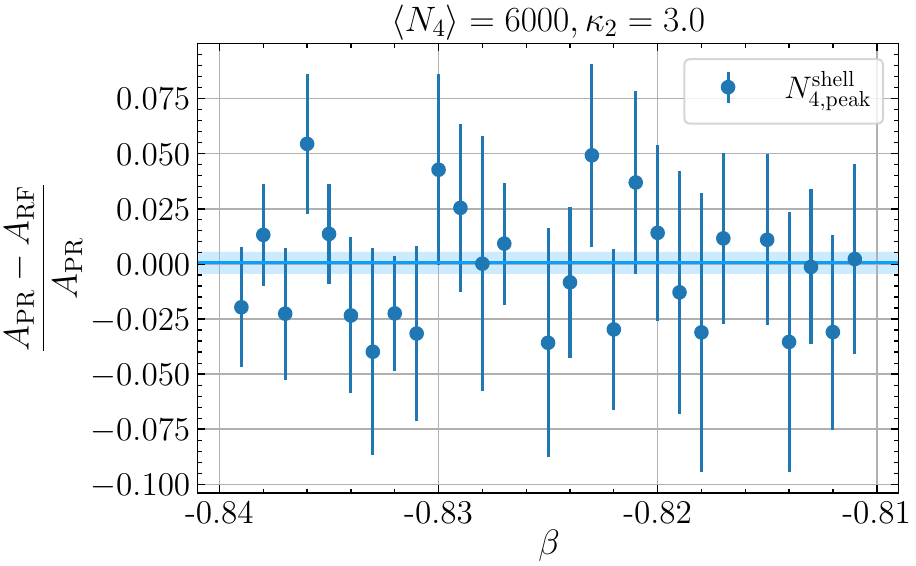}
    \end{minipage}
    \hfill
    \begin{minipage}{.495\linewidth}
    \includegraphics[width=1\linewidth]{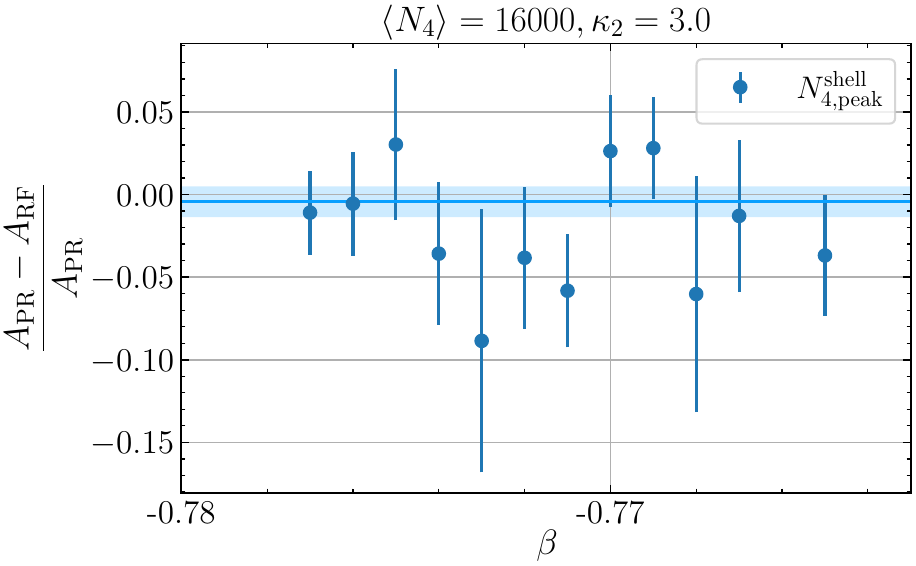}
    \end{minipage}
    \caption{Relative difference of the peak height $N_{4\,,\mathrm{peak}}^{\rm shell}$ of the shelling function $N_4^{\rm shell}(\tau)$ between the two algorithms. We show the results for fixed $\kappa_2$, and as a function of $\beta$. \newline
    Left panel: $\left\langle N_4 \right\rangle=6000$, and $\kappa_2=3$. The fit to a constant $C$ yields $C=0.0004(50)$ with $\chi^2_{\rm red}/\mathrm{d.o.f.}=0.523$. \newline 
    Right panel: $\left\langle N_4 \right\rangle=16000$, and $\kappa_2=3$. The fit to a constant $C$ yields $C=-0.004(9)$ with $\chi^2_{\rm red}/\mathrm{d.o.f.}=0.951$.}
    \label{fig:PeakHeightValid}
\end{figure*}

We introduce the shelling function $N_4^{\rm shell}(\tau)$ in order to characterize the emergent shape of our lattice geometries.  $N_4^{\rm shell}(\tau)$ is the total number of four-simplices in a spherical shell one four-simplex thick, a geodesic distance $\tau$ away from a randomly chosen four-simplex.  This function is averaged over multiple random sources on a given configuration and over all configurations of an ensemble.  The shelling function differs markedly in the different phases of EDT, so that the height of the peak of the shelling function $N_{4\,,\mathrm{peak}}^{\rm shell}$ serves as a good order parameter for studying the phase diagram.  We therefore use it for one of our validation tests by comparing the peak height results between the rejection free and standard Metropolis algorithms. \autoref{fig:PeakHeightValid} shows the relative difference of $N_{4\,,\mathrm{peak}}^{\rm shell}$ between the two algorithms at $\kappa_2=3.0$ for two different volumes. The relative difference is statistically compatible with zero for the full data set, where a constant fit to the difference gives $C=0.0004(50)$ for the left panel ($\left\langle N_4 \right\rangle=6000$), and a similar fit to the data in the right panel gives $C=-0.004(9)$ ($\left\langle N_4 \right\rangle=16000$).  The errors are dominated by the statistical error of the Metropolis algorithm runs, since it is more time consuming to take data using this slower algorithm.  The agreement serves as a useful check of the new algorithm. \\

\paragraph{Lattice observables\\[6pt]}

\begin{figure}
    \centering
    \includegraphics[width=\linewidth]{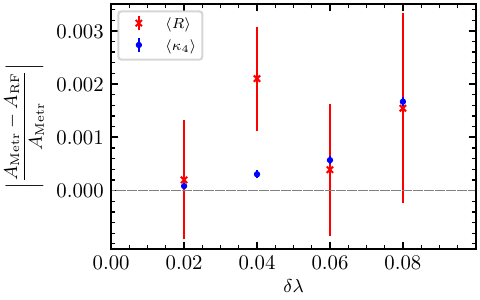}
    \caption{Relative difference for the average Regge curvature and average $\kappa_4$ between the metropolis and rejection free algorithm for fixed $\kappa_2=3.0$, $\beta=-0.822$, $\left\langle N_4 \right\rangle=6000$, as a function of the volume preserving term $\delta\lambda$. The relative difference decreases for both quantities with decreasing $\delta\lambda$, indicating that both algorithms approach agreement in the $\delta\lambda\to0$ limit.}
    \label{fig:EDTValid}
\end{figure}

As an even stronger test, we focus on a single ensemble and compare two lattice quantities that can be determined to higher precision.  We study the average Regge curvature \cite{Coumbe:2014nea}
\begin{equation}
    \left\langle R \right\rangle\approx \frac{2\,\pi}{10\,\arccos(1/4)}\left\langle\frac{N_2}{N_4}\right\rangle-1\,,
\end{equation} 
and the tuned $\left\langle \kappa_4 \right\rangle$ value.  \autoref{fig:EDTValid} shows the relative difference between these quantities for the two different algorithms. We display them as a function of $\delta\lambda$, the coefficient of the volume preserving term introduced in the simulations.  The Pachner moves that evolve the Markov chain are not ergotic, except in the limit that $\delta\lambda$ goes to zero, so a small systematic error is expected for all observables within the approximation of finite $\delta\lambda$.  Thus, a precision comparison of observables using different algorithms must take the $\delta\lambda\rightarrow 0$ limit.

As one can see in \autoref{fig:EDTValid} the two algorithms agree at the per-mil level, even for rather large values of $\delta\lambda$. Furthermore, the relative differences decrease with decreasing $\delta\lambda$, showing that in the limit $\delta\lambda\to0$ the results of the two algorithms agree.  \autoref{fig:EDTValid} also indicates that the residual error from not taking the $\delta\lambda\rightarrow 0$ limit is small. We have chosen $\delta\lambda=0.04$ for most of our simulations as a compromise between precision and performance, since the cost of simulations grows as $\delta\lambda$ shrinks.

The good agreement between the new rejection-free algorithm and standard Metropolis gives us confidence that the new algorithm is also behaving correctly for 4$d$ EDT.

\subsubsection{Performance gains}
Having established that the rejection free algorithm agrees with the Metropolis algorithm, we turn to measurements of the performance gains for EDT.  After each local update the rejection free algorithm requires that the ponderances associated with that local region be updated.  This is the main overhead associated with the rejection free algorithm, so to get a performance gain, this extra overhead must not take longer than the time saved from not having any proposed moves rejected.  In the $2d$ Ising model, there are only four neighbors to each spin, so with each spin flip only five ponderances need to be updated.  In $4d$ EDT the situation is more complicated.  The local measure term involves the order of triangles, i.e. the number of 4-simplices to which they belong.  When a local region of the geometry is updated, all of the ponderances associated with neighboring simplices and sub-simplices must be updated, and this can involve any simplex with a triangle that lives on the boundary of the local region.  The number of ponderance updates thus depends on the connectivity of the geometry, with high connectivity regions requiring a large number of updates.
This increase in overhead leads to a smaller speed-up at finer lattices, which seem to have regions of high connectivity.  

The red crosses in \autoref{fig:IsingPerformance} show the speed-up \eqref{eq:speedup} of the rejection free algorithm for EDT, compared to the previously used \textit{parallel-rejection} algorithm \cite{Laiho:2016nlp}. Both algorithms ran on four cores, and the same set of processors was used for all the runs shown in \autoref{fig:IsingPerformance}. For EDT, the acceptance rate changes when changing the bare parameters $\kappa_2$ and $\beta$, with the acceptance rate falling with larger $\kappa_2$ and more negative $\beta$. The displayed data points are taken to be close to what we have identified as the physical region of the phase diagram.

We see in \autoref{fig:IsingPerformance} that the rejection-free algorithm provides a speed-up of up to two orders of magnitude, which is achieved at an acceptance rate of about $10^{-5}$. The speed-up starts to plateau as one moves to even lower acceptance rates. The reason for this plateau lies in the increased connectivity of the local geometry as we move to larger values of $\kappa_2$. For the future this might pose a serious obstacle to performing efficient simulations at even larger $\kappa_2$; such simulations will likely require further optimization of the parallel ponderance updates to ameliorate this bottle-neck of the rejection free algorithm. For now, however, our implementation of the rejection-free algorithm provides us with about two orders of magnitude in speedup in the region of interest, and this allows us to investigate ensembles at finer lattice spacings and larger volumes than before.

\section{Results for EDT}\label{sec:Results}
We use the performance gains of our new algorithm to continue to study the EDT phase diagram in a region that was previously inaccessible to simulations.  We focus in particular on the region that earlier studies suggested would lead to finer lattice spacings \cite{Laiho:2016nlp}.  We find similar results to those studies, where a particular tuning of the bare parameters leads to an approximately four-dimensional, semi-classical Euclidean de Sitter geometry emerging from the simulations.  Our new results show that the agreement between our simulations and the classical limit improves at finer lattice spacings, continuing the trend found in Ref.~\cite{Laiho:2016nlp}.
\subsection{The phase diagram of EDT}

\begin{figure*}
    \centering
    \includegraphics[width=0.4\linewidth]{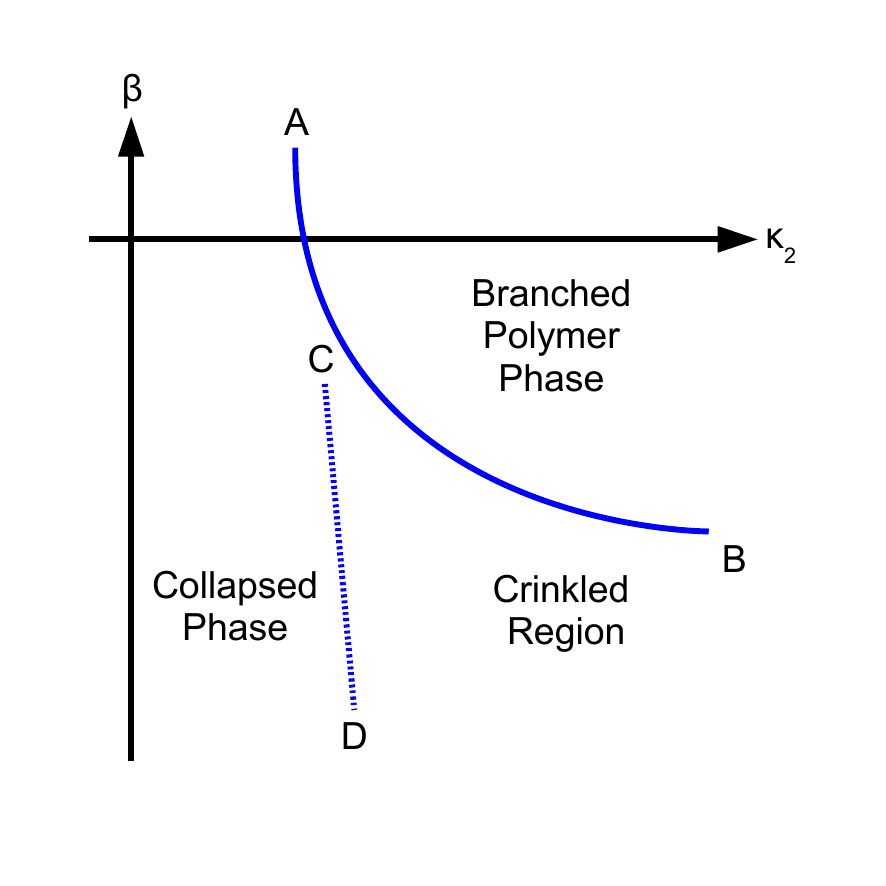}
    \caption{Schematic of the phase diagram as a function of $\kappa_2$ and $\beta$.}
    \label{fig:phaseDiagram}
\end{figure*}

The EDT phase diagram was first studied in the nineties \cite{Catterall:1994pg, Bialas:1996wu, deBakker:1996zx, Catterall:1997xj, Bilke:1998vj}, and also more recently \cite{Ambjorn:2013eha, Coumbe:2014nea, Rindlisbacher:2015ewa, Laiho:2016nlp, Asaduzzaman:2022kxz}.  The parameter $\kappa_4$ must be adjusted to set the lattice volume, leaving a two-dimensional parameter space in $\kappa_2$ and $\beta$ by which to explore the phase diagram.  A schematic of the phase diagram is shown in Fig. \ref{fig:phaseDiagram}. It was demonstrated already quite early on that there are two phases in the theory: a collapsed phase with large, possibly infinite, fractal dimension, and a branched polymer phase with dimension below four.  The solid line $\overline{AB}$ in Fig. \ref{fig:phaseDiagram} shows a phase transition separating the collapsed and branched polymer phases; there is now substantial evidence that this transition line is first order \cite{Bialas:1996wu, deBakker:1996zx, Coumbe:2014nea, Rindlisbacher:2015ewa}.  The crinkled region \cite{Bilke:1998vj} on the phase diagram shares features of both phases, but looks like the collapsed phase for sufficiently large volumes \cite{Ambjorn:2013eha, Coumbe:2014nea}. There does not appear to be a distinct phase transition between the crinkled region and the collapsed phase, but rather a crossover, as indicated by the dashed line CD in Fig \ref{fig:phaseDiagram}.

For generic bare coupling values the two phases do not bear much resemblance to a four-dimensional, semi-classical solution to the Einstein equations, and with a first order phase transition between them, an approach to that line cannot lead to the diverging correlation lengths needed to define a continuum limit.  However, the expanded diagram that is obtained when adding the local measure term suggests a way forward when we follow an analogy to lattice QCD with Wilson fermions, as first suggested in Ref. \cite{Laiho:2016nlp}. In this picture, one must tune close to the first-order line in order to recover semiclassical physics, and follow this line out to a possible critical end-point.

In the study of a UV-fixed point for gravity we do not have the benefit of being able to use perturbation theory at short distance scales, as in the case of QCD, which is asymptotically free.  In the EDT formulation, we are more dependent on numerical simulations to investigate the short-distance behavior.  There are two tests that EDT must pass in order to realize the asymptotic safety scenario for gravity.  First, there must be a continuous phase transition, where the associated diverging correlation length would allow the lattice spacing to be taken to zero.  Second, the formulation must recover the classical Einstein theory in four dimensions, since this is a good description of our world.  Refs. \cite{Laiho:2016nlp, Catterall:2018dns,  Dai:2021fqb, Bassler:2021pzt, Asaduzzaman:2022kxz} provide evidence in favor of these tests, but further work is necessary.  Because the Metropolis acceptance rate drops rapidly as we take $\kappa_2$ large, it is difficult to follow the phase transition line in Fig. \ref{fig:phaseDiagram} out to large $\kappa_2$.  The rejection free algorithm introduced here allows us to push further into this regime and to generate ensembles that were not feasible with previous algorithms.  In the rest of this section we present a first look at the basic geometric properties of these new ensembles. 

This section revisits the shelling function $N_4^{\rm shell}(\tau)$ introduced in Section~\ref{sec: Tests} as part of our validation tests. Here it is used to characterize the emergent shape of our lattice geometries.  Given its sharp dependence on the phase one is simulating, the height of the peak of the shelling function serves as a good order parameter for identifying the location of the phase transition.  It is convenient to consider the rescaled shelling function 
\bea n_4(\rho) = \frac{1}{N_4^{1-1/D_H}}N_4^{\rm shell}(N_4^{1/D_H}\rho),
\eea
where $\rho=\tau/N_4^{1/D_H}$ is the rescaled Euclidean distance, and $D_H$ is the Hausdorff dimension.  The Hausdorff dimension is a fractal dimension that is defined by the scaling of the volume of a sphere with its radius.  

\begin{figure*}
    \begin{minipage}{0.495\linewidth}
    \centering
    \includegraphics[width=1\linewidth]{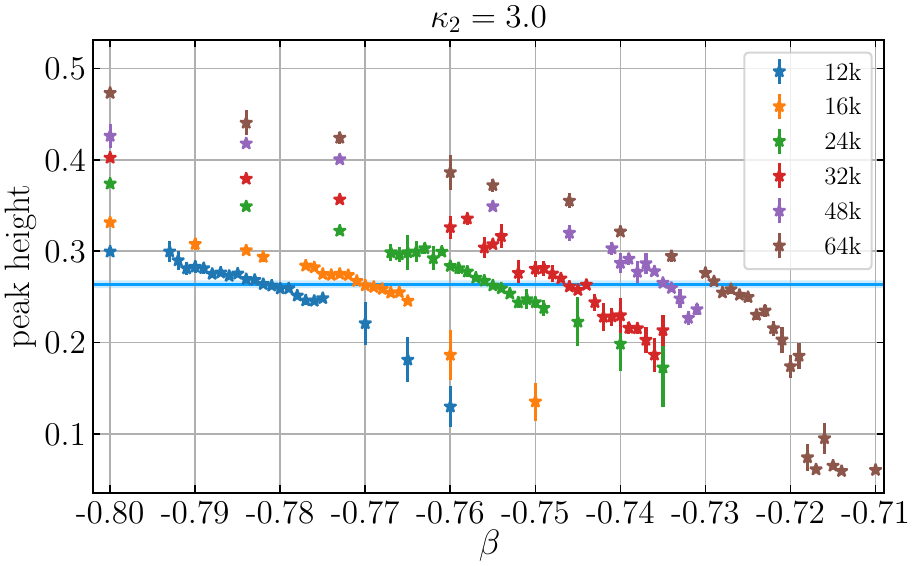}
    \end{minipage}
    \begin{minipage}{0.495\linewidth}
    \centering
    \includegraphics[width=1\linewidth]{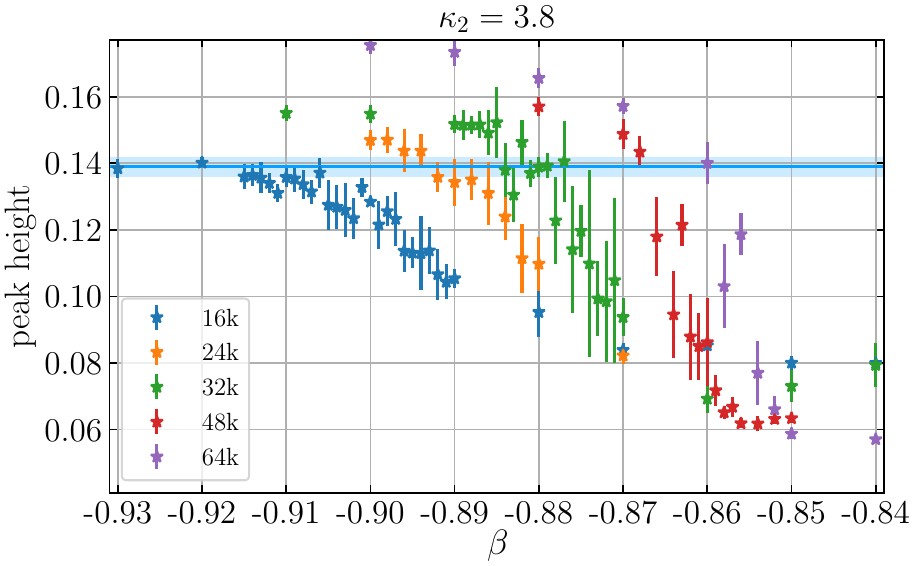}
    \end{minipage}
    \caption{Left panel: Peak height of shelling function versus $\beta$ at several volumes for $\kappa_2 = 3.0$ ensembles. The cyan band represents the peak height that is used to match against for the different volumes at the same nominal lattice spacing.    \newline
    Right panel: Same as left panel but for $\kappa_2 = 3.8$.  Note that the cyan band is close to the ``knee'' just before the slope in the lattice data becomes large and negative, independent of the volume. }
    \label{fig:f&sfTuning}
\end{figure*}

Figure \ref{fig:f&sfTuning} shows a plot of peak height of the rescaled shelling curve $n_4(\rho)$ as a function of $\beta$ with $\kappa_2$ fixed.  For the rescaling we choose $D_H=4$.  In Fig. \ref{fig:f&sfTuning}, the phase to the left with more negative beta is the collapsed phase, and the phase to the right is the branched polymer phase.  There is evidence for a phase transition in between, where the decrease in peak height as a function of $\beta$ becomes more pronounced as the volume is increased.  There is a ``knee'' in the plots at each volume, just before the slope becomes large and negative, that marks the onset of the phase transition.  The fact that this knee occurs at roughly the same peak height in the rescaled volume is evidence that the lattice geometries have Hausdorff dimension close to four near the phase transition, since the rescaling was done with $D_H=4$.  The left panel of Fig. \ref{fig:f&sfTuning} shows data at the finest lattice spacing that was generated in previous work \cite{Laiho:2016nlp}, where the simulations were restricted to the smallest volumes.  The right plot shows an even finer lattice spacing that was infeasible to simulate with earlier algorithms and is new to this work.  The cyan band in each plot is the peak height of the $n_4(\rho)$ function that has been chosen as the tuned value for that lattice spacing.  The ensembles at each volume at a fixed lattice spacing (fixed $\kappa_2$ value) are chosen to match this value of the peak height as closely as possible.  In both the right and left panels of Fig.~\ref{fig:f&sfTuning} it can be seen that as the volume is increased, the curve becomes steeper, typical behavior for a phase transition.  We have not seen the tunneling between meta-stable states that is characteristic of a first order phase transition at these larger values of $\kappa_2$, but it may be that we have not yet reached large enough volumes for the effect to be visible.

\begin{table*}[]
\begin{tabular}{|c|c|c|c|c|c|c|c||c|c|c|c|c|}
\multicolumn{8}{c}{$\kappa_2=3.0$}                                                                                 & \multicolumn{5}{c}{$\kappa_2 = 3.8$}       \\ \hline
Volume ($N_4$)  & 8k  & 12k   & 16k & 24k & 32k  & 48k  & 64k  & 16k & 24k  & 32k & 48k & 64k \\ \hline
$\beta$     & -0.80        & -0.782 & -0.771                     & -0.756                     & -0.746 & -0.735 & -0.729 & -0.92 & -0.894 & -0.88 & -0.868 & -0.86 \\\hline
\#configurations & 1486 & 17350  & 21114                      & 30555                      & 36142  & 24295  & 97030  & 6096  & 2676   & 13137 & 5714  & 10096\\ \hline
Block size & 743 & 946 & 782 & 507 & 1280 & 1052 & 998 & 879 & 578 & 517 & 986 & 1053 \\ \hline
\end{tabular}

\caption{$\beta$ and number of configurations used in analysis of tuned ensembles of $\kappa_2=3.0$ and $\kappa_2=3.8$.}
\label{tab:tunedensembles}
\end{table*}

\begin{figure*}
    \begin{minipage}{0.495\linewidth}
    \centering
    \includegraphics[width=1\linewidth]{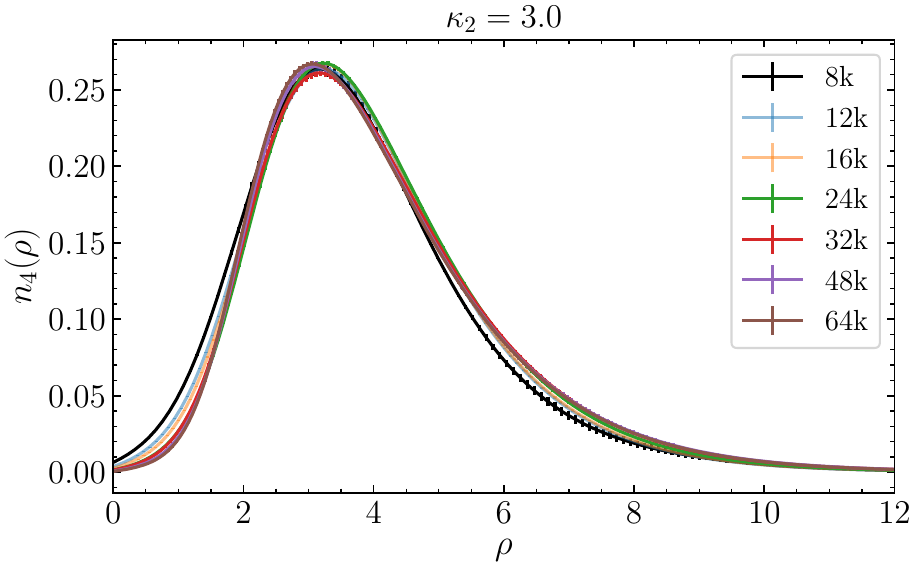}
    \end{minipage}
    \begin{minipage}{0.495\linewidth}
    \centering
    \includegraphics[width=1\linewidth]{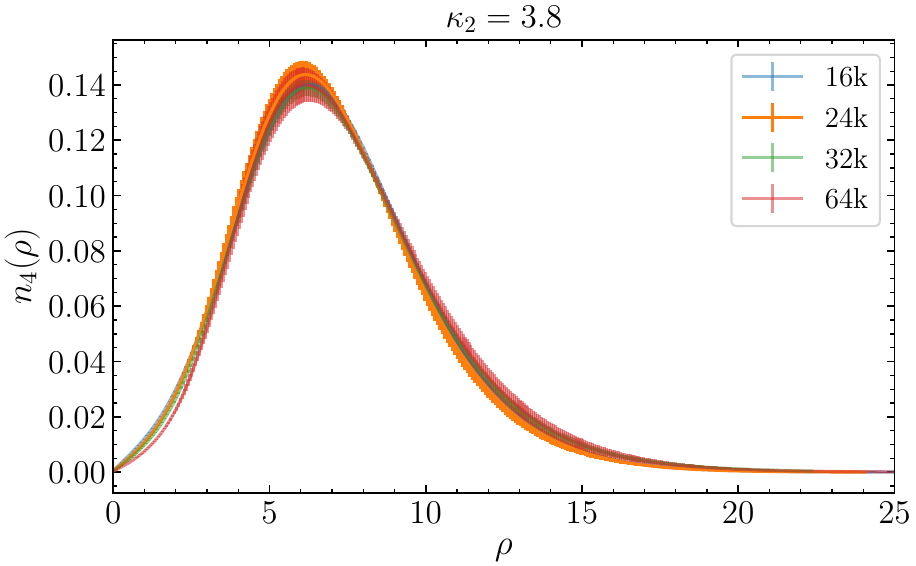}
    \end{minipage}
    \caption{Left panel: The rescaled shelling function $n_4(\rho)$ at the tuned $\beta$ values for a number of volumes at the same nominal lattice spacing, with $\kappa_2 = 3.0$.  The rescaling was done assuming $D_H = 4$ for all ensembles.\newline Right panel: The rescaled shelling function $n_4(\rho)$ for a finer lattice spacing, with $\kappa_2 = 3.8$, again assuming $D_H = 4$.  With $D_H=4$, the horizontal axis and vertical axis have been rescaled by $N_4^{1/4}$ and $N_4^{3/4}$, respectively.}
    \label{fig:4dscaling}
\end{figure*}

The tuned value of $\beta$ at a fixed nominal lattice spacing (fixed value of $\kappa_2$) is determined by matching the peak heights in the vicinity of the phase transition.  A tuning to the ``knee'' just to the left of the phase transition gives a $D_H$ close to four. For convenience we assume that the peak height rescaling is four-dimensional, and we tune all of the other volume ensembles to the position of the ``knee'' on one of our smaller volumes, the 8k volume at $\kappa_2=3.0$ and to the 32k at $\kappa_2=3.8$.  The agreement in Fig.~\ref{fig:f&sfTuning} between the cyan band and the location of the ``knee'' across volumes is an indication of the $D_H=4$ nature of the geometries in the vicinity of the phase transition.  
 
 Figure~\ref{fig:4dscaling} shows the shelling functions $n_4(\rho)$ at the tuned $\beta$ values with $D_H=4$ chosen for the rescaling.  To compute the shelling function, 60 sources are chosen randomly on each configuration.  We account for auto-correlation errors present in the data for the shelling function by blocking the data before averaging.  We study the variation of the error with block size, and we continue to block until the error no longer increases after taking the jackknife average over configurations and computing its standard error.  Table \ref{tab:tunedensembles} shows the parameters of the tuned ensembles and the number of configurations used to calculate the shelling function on each ensemble, as well as the block-size needed to account for autocorrelation errors.  In addition to the standard rescaling of the horizontal and vertical axes, a small shift in the horizontal direction that is proportional to the inverse of the volume is applied in order to make the location of the peaks coincide on the plot. This shift is simply a knob which allows the collapse of the shelling function in the semi-classical regime, and does not change the shape of the curve.  We expect at very short distances that lattice artifacts distort the volume distribution, and at slightly longer distances, possibly unknown short-range physics dominates.

The agreement over the entire shelling curve is not perfect, though the differences are not very big in absolute terms. Since the peak was matched across volumes assuming $D_H=4$ at a given nominal lattice spacing (fixed $\kappa_2$ value), the agreement of the peak in these curves is assumed, but the agreement of the full shape of the shelling function across volumes is non-trivial.  The matching of the peak heights to $D_H=4$ is not arbitrary, but follows from the good agreement of the knee just to the left of the phase transition (see, for example the right panel of Fig.~\ref{fig:f&sfTuning}) with the cyan band.  We conclude that the large-scale geometry of our lattices is consistent with $D_{H}=4$, which agrees with the desired behavior. 

\subsection{Approaching the de Sitter solution}
\begin{figure}[ht]
    \centering
    \includegraphics[width=1\linewidth]{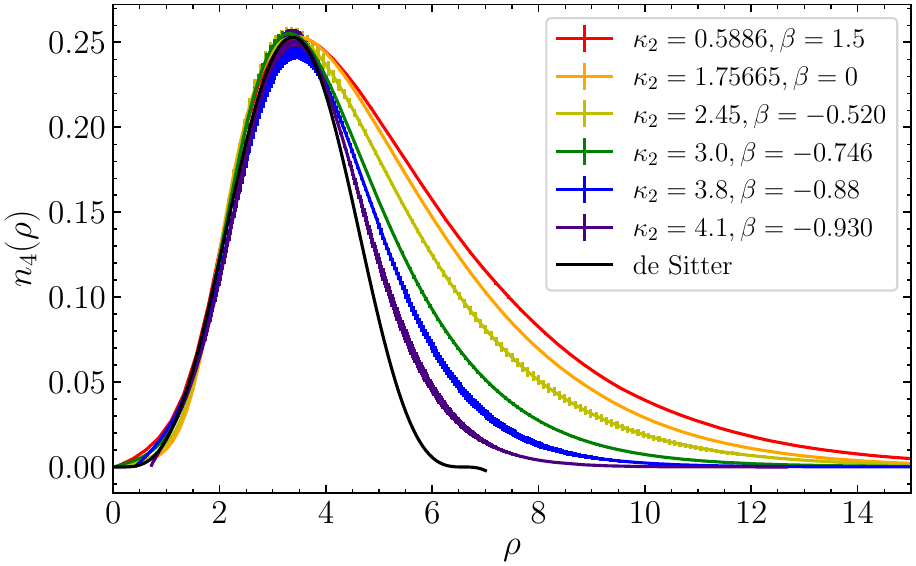}
    \caption{The rescaled shelling function $n_4(\rho)$ at different lattice spacings.  The black curve is the classical de Sitter solution.  All curves have been rescaled to overlap in the region that matches the classical solution.  The asymmetry at large Euclidean time decreases as the lattice spacing gets finer, so that the lattice results approach the classical curve.}
    \label{fig:deSitteraAymTail}
\end{figure}

Since the shelling function describes the emergent shape of the lattice, it is natural to associate it with the global dynamics of the scale factor of the universe and to try to match the resulting dynamics to a cosmological solution of the Einstein equations.  Previous work \cite{Laiho:2016nlp} identified this with Euclidean de Sitter space, which is the solution to the (Euclidean) Einstein equations with a positive cosmological constant.  The Euclidean de Sitter solution can be expressed as 
\bea\label{eq:deSitter}  N_4^{\rm shell} = \frac{3}{4}N_4 \frac{1}{s_0 N_4^{\frac{1}{4}}}\sin^3\left(\frac{j}{s_0 N_4^{\frac{1}{4}}}\right),
\eea
where the exponential expansion of the de Sitter solution in real time has transformed to an oscillating function after continuation.  In this expression, $s_0$ is a free parameter and $j$ is the Euclidean time in lattice units.  In order to describe the lattice data, we need to add some free parameters to this formula,  
\bea\label{eq:deSitterMod}  N_4^{\rm shell} = \frac{3}{4}\eta N_4 \frac{1}{s_0 N_4^{\frac{1}{4}}}\sin^3\left(\frac{j}{s_0 N_4^{\frac{1}{4}}} + b\right),
\eea
where we have introduced the parameters $b$ and $\eta$. The parameter $b$ is an offset in the Euclidean time, and $\eta$ accounts for the volume of the universe that is actually well-described by the classical solution.  Figure~\ref{fig:deSitteraAymTail} shows a comparison between the theoretical expectation, Eq.~(\ref{eq:deSitterMod}), and the data at multiple lattice spacings.  The lattice data has been rescaled so that each data set overlaps with the classical de Sitter solution in the region to the left of the classical peak.  It is then clear that there is a large discrepancy between the classical solution and the data at large Euclidean time, but this discrepancy gets smaller as one takes the lattice spacing ever finer.  This trend was first demonstrated in Ref.~\cite{Laiho:2016nlp}.  The present work follows this up by adding two additional finer lattice spacings, and Fig.~\ref{fig:deSitteraAymTail} shows that the trend towards greater agreement continues as we follow $\kappa_2$ to larger values.  Work in progress shows that the relative lattice spacing also continues to decrease as we take $\kappa_2$ larger, so that improved agreement with the classical curve occurs at finer lattice spacings.  

Note that the biggest discrepancy here occurs at large Euclidean time, and although one might expect that long-distance quantities should not be modified by lattice cutoff effects, this can happen when the lattice regulator breaks a symmetry that is important at long distances, such as chiral symmetry in the case of Wilson fermions. The pion sector of QCD is not correctly described at coarse lattice spacings for Wilson fermions, even though this is the lightest particle in the physical spectrum with the longest Compton wavelength. Only in the continuum limit does the residual chiral symmetry breaking vanish \cite{Bernard:2004ab}.  The behavior seen here is reminiscent of that of QCD with Wilson fermions, further suggesting that a symmetry that is broken by the EDT lattice regulator is restored in the continuum limit.  This is one of the main arguments suggesting that the tuning of $\beta$ is needed to restore a broken symmetry of EDT.  If the symmetry is an exact symmetry of the quantum theory, as one would expect if the symmetry is continuum diffeomorphism invariance, then the number of relevant parameters in the symmetry-preserving theory would be less than the three that are needed in the lattice formulation.  Reconciling this result with the three-dimensional ultra-violet critical surface found in many different truncations of the FRG is a strong motivation for future work.

\section{Conclusion and Outlook}\label{sec: Concl}

In this work we have introduced a new, faster algorithm for doing simulations of EDT.  This algorithm mimics the Metropolis algorithm in its evolution along the Markov chain, but its proposals are never rejected.  Such algorithms have been around for a long time, but this one is new to the best of our knowledge, in that it allows for the factorization of local and global terms in the update while still maintaining detailed balance.  This is made possible by introducing what we call the ponderance, which is the square root of the standard Metropolis accept probability.  The rejection-free algorithm requires keeping track of the ponderances of all possible moves as the Markov chain progresses.  After each local move, all possible moves whose associated ponderance is altered must be updated. The presence of global terms in the action that change after every local move requires the factorization of global and local terms for an efficient implementation.  Otherwise, all probabilities would need to be updated, which would be prohibitively expensive.  By using the ponderance instead of the probability, we have shown that it is possible to factor the action into local and global parts, leading to an efficient rejection-free algorithm for EDT.

We test this algorithm on the $2d$ Ising model, where we show that the results from the new algorithm match those of the Metropolis algorithm, and they also match the analytical results in the low temperature regime.  This validation gives us confidence that the algorithm is behaving as it should.  This algorithm has significant advantages when the Metropolis acceptance rate is low, and as expected, the performance is orders of magnitude better for the Ising model in the low-temperature regime, where the standard Metropolis acceptance rate is low.  We further test the new rejection free algorithm on $4d$ EDT, where we once again see good agreement between rejection free and standard Metropolis.  A similar improvement in performance is seen for EDT, which also suffers from the low acceptance of standard Metropolis, especially as one moves to finer lattice spacings.  The improvement of EDT is not as pronounced as that of the Ising model at very low acceptances because of the high connectivity of the EDT model, at least in the region of the phase diagram that we explore.  This could potentially be improved by additional parallelization of the ponderance updates after each move.

Our new algorithm allows us to simulate in a region of the phase diagram of EDT that was previously inaccessible, and thus to further test whether the results match expectations if EDT is to provide a viable formulation of quantum gravity.  The qualitative behavior of the phase transition that was observed in Ref.~\cite{Laiho:2016nlp} does persist out to larger values of the coupling $\kappa_2$, with the emergence of semi-classical geometries that approximately match the Euclidean de Sitter solution close to the phase transition.  The agreement with the classical solution gets better as the lattice spacing is made finer, and there appears to be no barrier, in principle, to taking the continuum limit.  A more detailed study of the new EDT ensembles that were generated with the rejection-free algorithm introduced here will be presented in forthcoming work.

\begin{acknowledgments}
The authors thank Gwen Hartshaw for valuable discussions, and we thank Simon Catterall for comments on the manuscript.  JL was supported by the U.S. Department of Energy (DOE), Office of Science, Office of High Energy
Physics under Award Number DE-SC0009998.  JUY was supported by the U.S. Department of Energy grant DE-SC0019139, and by Fermi
Research Alliance, LLC under Contract No. DE-AC02-
07CH11359 with the U.S. Department of Energy, Office
of Science, Office of High Energy Physics. MS acknowledges support by Perimeter Institute for Theoretical Physics. Research at Perimeter Institute is supported in part by the Government of Canada through the Department of Innovation, Science and Economic Development and by the Province of Ontario through the Ministry of Colleges and Universities. Computations were performed in part on the Syracuse University HTC Campus Grid and were supported by NSF award ACI-1341006. This research was enabled in part by support provided by ACENET, Calcul Québec, Compute Ontario, and the Digital Research Alliance of Canada. The authors acknowledge support by the state of Baden-Württemberg through bwHPC.
\end{acknowledgments}

\appendix
\section{Recovering Monte-Carlo time from rejection free algorithms}
\label{app:nreject}

This appendix discusses a rejection-free algorithm that does not have a global contribution to the action.  In that case we emulate the Metropolis algorithm directly.

In short, we need to determine:

\begin{itemize}
\item Which move $i$ will be the first one to be accepted by Metropolis
\item How many tries it took to accept that move (since per the Metropolis
algorithm simulation time is measured in attempted moves, not accepted ones)
\end{itemize}

Call the probability that any given move $i$ will be the one eventually
accepted $\tilde P(i)$. Then

\begin{equation}
\tilde P(i) = \frac{P(i)}{\sum_j P(j)},
\end{equation}
that is, the probability of eventually accepting any particular move $i$ is a fraction of the
sum over Metropolis probabilities for all possible successive moves.

For the second point, the number of Metropolis suggestions that are rejected before one is eventually
accepted is independent of which one is eventually chosen. Thus it suffices to
determine the probability that any given Metropolis suggestion will result in a rejection.

The probability of accepting any move $i$ on any particular trial
is equal to the probability of choosing it
multiplied by the probability of accepting it, {\it i.e.}

\begin{equation}
P_a(i) = \frac{P(i)}{N}
\end{equation}
where $N$ is the total number of possible moves that could be chosen.

Since these probabilities are mutually exclusive, we can just add them, to determine that
the probability of accepting the move that we choose is

\begin{equation}
P_a = \frac{\sum_i P(i)}{N}
\end{equation}
that is, the probability that the move that is chosen at random will be accepted is just the average accept probability of all the moves.

However, the Metropolis Markov chain will have a long sequence of one configuration, since every time a move is rejected, that configuration
is added to the chain again. It is thus insufficient to calculate the next configuration; one must also calculate how long the system
stays stuck in the current one.

The probability of accepting any move after rejecting $n$ previous trials is thus

\begin{equation}
P(n) = P_a (1-P_a)^n
\end{equation}
and the number of previously rejected trials can be determined from a single random number $r$ from (0,1) as

\begin{equation}
    n_{\text{reject}} = \text{floor}(\log_{1-P_a}(r)).
    \label{eq:nreject}
\end{equation}

\section{Implementation details of a binary decision tree}
\label{app:bintrees}

The rejection-free algorithm requires us to select a move from amongst a list of $N$ moves such that the relative probability of selecting any given move $i$ is proportional to its local ponderance $\mathcal P_{\rm{loc}}(i)$. This can be accomplished in $\mathcal O(\log N)$ time using a binary decision tree. 

Such a tree contains one node for each possible move. Each node in the tree has two child nodes, such that storing $N$ moves in the tree requires $\log_2 N$ layers. Each node in the tree is indexed by a move index $i$; this index need have no relation to the geometry or properties of the physics being simulated.

Each node contains three pieces of data:

\begin{enumerate}
\item The move index $i$
\item The local contribution to the ponderance of that move $\mathcal P_{\rm{loc}}(i)$, which we call $\mathcal{P}_{\rm here}$
\item The sum of the values of $\mathcal P_{\rm{loc}}(i)$ of this node and all of its children, which we call $\mathcal{P}_{\rm below}$
\end{enumerate}

To determine which move $j$ will be the one that the Metropolis algorithm eventually accepts:

\begin{enumerate}
\item Generate a random number $r$ between 0 and the  sum of all $\left(\mathcal{P}(i)\right)_{\mathrm{loc}}$ in the whole tree (which is equal to $\mathcal{P}_{\rm below}$ in the root node).
\item Start at the root of the tree and traverse it, looking for the node $j$ that will be accepted.
\item At each node, we have three options: the move we are seeking either is the one in the current node, or it lives somewhere along the left or the right branch.
\item Define $\mathcal{P}_{\rm here}$ as the probability of the current node, $\mathcal{P}_{\rm left}$ as the total probability along the left branch, and $\mathcal{P}_{\rm right}$ as the total probability along the right branch. (These quantities are stored as $\mathcal{P}_{\rm below}$ in the child nodes so they are quick to access.)
\item Traverse the tree according to the following:
\begin{itemize}
\item If  $r < \mathcal{P}_{\rm left}$, then go left and repeat the procedure
\item If  $\mathcal{P}_{\rm left}<r<\mathcal{P}_{\rm left} + \mathcal{P}_{\rm here}$, then the desired move $j$ is the one at the current node
\item If $r>\mathcal{P}_{\rm left} + \mathcal{P}_{\rm here}$, then go right and redefine $r \rightarrow r-(\mathcal{P}_{\rm left} + \mathcal{P}_{\rm here})$ and repeat the procedure
\end{itemize}
\end{enumerate}

\bibliographystyle{apsrev4-1}
\bibliography{references.bib}
\end{document}